# Reproducible experiments on dynamic resource allocation in cloud data centers

Andreas Wolke

Technische Universität München, Boltzmannstraße 3, 85748 Garching, Germany
wolke@tum.de

Martin Bichler

Technische Universität München, Boltzmannstraße 3, 85748 Garching, Germany
bichler@in.tum.de

Fernando Chirigati[1]

New York University, United States
fchirigati@nyu.edu

Victoria Steeves[1]

New York University, United States
vs77@nyu.edu

**Refers To**
Andreas Wolke, Boldbaatar Tsend-Ayush, Carl Pfeiffer, Martin Bichler, More than bin packing: Dynamic resource allocation strategies in cloud data centers, Information Systems, Volume 52, August–September 2015, Pages 83-95 | PDF (622 K)

## Abstract

In Wolke et al. [1] we compare the efficiency of different resource allocation strategies experimentally. We focused on dynamic environments where virtual machines need to be allocated and deallocated to servers over time. In this companion paper, we describe the simulation framework and how to run simulations to replicate experiments or run new experiments within the framework. Data accompanying this article can be found here: https://data.mendeley.com/datasets/xz6gv65m6d/6

## 1. Introduction

Reproducibility is the ability of an entire experiment or study to be duplicated, and it constitutes one of the main principles of the scientific method. Research on resource allocation in cloud computing largely consists of discrete event simulations. The results are difficult to replicate and hard to compare as they are typically based on different assumptions and implementations. This not only hinders progress, but it does also not allow for reliable results expected by academics, practitioners, and other interested parties. Twenty years ago Tichy et al. [2] criticized that "the low ratio of validated results appears to be a serious weakness in computer science research."

Many academic fields have developed standards to ensure reproducibility of their experiments. For example, microeconomists developed strict guidelines on how to conduct and report experiments, which led to reliable empirical results about human behavior in economic interactions [3]. A challenge in microeconomics is the control of human subjects in a lab. In the systems literature, a challenge is the number of hard- and software components involved and the rapid technical progress of these. As in any other field of science and engineering it is still important that results can be reproduced by others and that the assumptions and details of the implementations are easily available to others. This does not only increase the credibility of the research, but it is also vital for the progress of a field.

Technology nowadays makes it possible to reveal not only the data for an experiment, but to also make it easier that others can access the simulation software such that they can reproduce the results. This article is a companion paper to [1], which reports the results of experiments on resource allocation algorithms for cloud computing infrastructures. In this companion paper, we describe the simulation software that is made available via a Docker container. We recommend readers to first read through the general experimental environment outlined in Wolke et al. [1], before reading through this paper.

In what follows, we will briefly revisit the results from [1] in the next section, describe the simulation framework, and how to run simulations in the framework. Finally, we will outline software dependencies relevant for replication experiments.

## 2. Experiments in Wolke et al. [1]

In Wolke et al. [1] we compare the efficiency of different resource allocation strategies experimentally. We focused on dynamic environments where virtual machines need to be allocated and deallocated to servers over time. Simple bin packing heuristics were analyzed and used to place virtual machines upon arrival. These placement heuristics can lead to suboptimal server utilization, because they cannot consider virtual machines, which arrive in the future.

We ran lab experiments and simulations with different controllers and different workloads to understand which control strategies achieve high levels of energy efficiency in different workload environments. Combinations of placement controllers and periodic reallocations achieved the highest energy efficiency subject to predefined service levels.

While the type of placement heuristic had little impact on the average server demand, the type of virtual machine resource demand estimate used for the placement decisions had a significant impact on the overall energy efficiency.

These results were generated using a software framework described in this article. The same software components and implementations of different control strategies were used for lab experiments and simulations, which required the development of a new software framework. We designed our software such that other researchers can extend it, implement their own resource allocation controllers, and benchmark them against existing controllers.

## 3. The simulation framework

A simulation is given a set of time series as well as the server and virtual machine (VM) capacity as an input. Each of the time series describes the CPU utilization of a VM. A server's utilization is described by the sum of the VM utilizations running plus the base demand of the server itself.

An *initial placement controller* is executed in the first step of a simulation. It computes a mapping of VMs to servers which is called VM allocation. Afterwards the VMs are migrated to the server accordingly. Subsequently, the simulation loop is initiated by injecting a message to the global message pump. At the end of a simulation loop, a new message is injected to trigger the next simulation loop within 3 s. Server and VM utilization levels are updated in each simulation loop.

A *reallocation controller* is triggered by the message pump in regular intervals, potentially triggering VM migrations. Usually, these take longer than the 3 s simulation loop interval and are controlled by the message pump as well. Migrations increase the CPU and memory utilization by variable amounts on the servers involved (both migration source and target server). Appropriate values are added during the server and VM load computation.

*Placement controllers* allow the simulation of dynamic cloud environments where VMs are allocated and removed continuously. For a simulation, this process is described by a VM arrival–departure schedule. For each VM allocation the placement controller is executed to determine a target server for the VM. At the end of a VM's lifetime it is removed from the simulation automatically.

The simulation framework mimics the CPU and memory utilization of servers and VMs in a cloud infrastructure. Neither applications running within the VMs are modeled nor the network infrastructure. In lab experiments, real VMs and hardware components can be used instead of the simulations. The implementation leveraging a physical server infrastructure depends on micro-services (e.g., a monitoring service), which are not described in this article. Wolke [4, Appendix A] provides an overview of the experimental testbed infrastructure we used and [4, Appendix C] explains how this infrastructure was controlled by the simulation framework.

All controller implementations presented in [1] are found in the Docker container folder `SRC_BALANCER=/root/work/paper.IS2015/control/Control/src/balancer` . Initial placement controllers extend from *InitialPlacement*, reallocation controllers from strategy. *StrategyBase*, and placement controllers extend placement. *PlacementBase* classes. For new controller implementations, a class/name mapping has to be added to the `SRC_BALANCER/controller.py` script.

## 4. Configuring a simulation

- No special configuration files exist in our framework as everything is configured within a set of Python source files in the folder `SRC_CTRL=/root/work/paper.IS2015/control/Control` with the filename prefix `conf_`.
- `SRC_CTRL/src/conf_controller.py` specifies which controllers to use for initial placement, reallocation, and placement controllers. Each value is a string that is mapped to a concrete class by the script `SRC_CTRL/src/balancer/controller.py` . Each controller requires a mapping within this script. An initial placement controller is required by all simulations while reallocation and placement controllers are optional. Controllers are disabled by setting their configuration value to *None*.
- `SRC_CTRL/src/conf_domainsize.py` contains the available domain[2] sizes with CPU and memory capacity and the probability that a domain size appears in a simulation.
- `SRC_CTRL/src/conf_domains.py` describes the number and capacity of each domain within the simulation. The setup is done procedurally and can be changed accordingly. It should be noted that the configured domains correspond to the state of a physical infrastructure if experiments are conducted.
- `SRC_CTRL/src/conf_load.py` holds a list of time series used during the simulation. Time series are stored in a special service called Times, described below, where they can be accessed by simulations as well as other services required by an experiment.
- `SRC_CTRL/src/conf_schedule.py` describes the VM arrival/departure schedule that is used to run simulations involving a placement controller. The *SCHEDULE_ID* refers to a JSON schedule configuration file within the directory `schedules`. These pre-configured schedules were used in our paper [1] but new ones can be re-generated using `SRC_CTRL/src/schedule/schedulebuilder.py` .
- `SRC_CTRL/src/configuration.py` running experiments are far more complicated than running simulations. The framework was designed in a way that allows users to conduct experiments by flipping a single configuration flag without detailed knowledge about the testbed infrastructure. This is what this central configuration file is for. It holds all relevant IP addresses and hostnames to control the testbed infrastructure. Technically, all controllers operate on APIs to control any infrastructure. In production mode these APIs are satisfied by implementations that control a real testbed infrastructure instead of a simulated one. This is achieved through numerous micro-services such as Times or Rain[3] and a well-defined operating system installation image as well as network configuration as described in [4, Appendix A]. This approach allowed us to carefully prepare experiments offline in a fast trial and error development cycle and eliminate almost all bugs before launching time-consuming experiments.

Time series are managed by a network service called Times.[4] It stores time series as binary blobs using the Apache Thrift [5] serialization and leverages Thrift to provide a network service where simulations and experiments can download time series.

## 5. Running simulations as a developer

Simulations can be started by running the Python script `SRC_CTRL/src/balancer/simulation.py`. For this, it is necessary to configure the IP address and port of the Times service upon installation. Core metrics such as the average server demand and CPU utilization are written to the standard output stream at the end of a simulation and optionally to a CSV file in the system's temp directory.

The script `SRC_CTRL/src/balancer/simulation_loop.py` is used to conduct a larger set of simulations with different parameters based on a factorlevel matrix. The factorlevel matrix is auto-generated based on a list of initial placement, reallocation, and placement controllers. Different configuration combinations already exist and are configured at the end of `SRC_CTRL/src/balancer/simulation_loop.py`. For each combination in the factorlevel matrix the script `SRC_CTRL/src/balancer/simulation.py` is executed in a child-process. Multiple child-processes are started in parallel to speed up processing.

The simulation_loop.py script has an output file name as a single argument. Two files with the given file name are written to the system's temp directory, one CSV with successful simulation results and an ERR file with failed simulation configurations (e.g., this happens if a mixed-integer program was intractable within the given time limits). To prevent existing files from being overwritten they are moved and suffixed with `.bak`.

After running a larger set of simulations, result data is aggregated by analysis scripts. The R Markdown file `SRC_ALS/analysis - simulations.Rmd` in the folder `SRC_ALS=/root/work/paper.IS2015/analysis` triggers the analysis process and generates all illustrations and tables as presented in [1].

## 6. Running simulations with Docker

After installing the Docker container (IS2015) it should be launched in interactive mode: *docker run -i -t IS2015 bash*. A full simulation run is started by `SRC_CTRL/startsim_reprozip` . The script will launch the Times service and trigger simulations as described in [Section 5](). Simulation results are written to /tmp/controller_sim.csv, failed simulations to `/tmp/controller_sim.err` (some failures are OK). An existing simulation output file is moved to a backup file `/tmp/controller_sim.csv.bak` , overwriting existing backup files.

Simulation results are analyzed by `SRC_ALS/startanalysis-sim` . It will generate an HTML output file `analysis_-_simulations.html` with all relevant tables and plots.

Simulations will run on any hardware as long as the required dependencies are available. However, results will slightly vary as some simulations solve mixed-integer problems with an upper bound on execution time. Their solution quality depends on the underlying machine. On a Core I7 870 CPU with 8 GByte of RAM the Docker container finished all simulations with 3 h.

## 7. Research artifacts

We provide a Docker container on Mendeley Data [6] that allows a full reproduction of our simulation results. In addition, the source codes and simulation input data are available as GitHub repositories:
1. Docker container with preinstalled simulation framework, source codes, and workloads [6].
2. GitHub repository[5] with source codes of the simulation framework.
3. GitHub repository[6] with workloads (time series data in Times format).

The following dependencies are required by the simulation framework. Alternatively the Docker container can be used to run the simulations without going through the installation process. A Dockerfile describing the whole installation process is available online [6].

1. Linux Fedora 21 with Development Tools and Development Libraries, Pandoc 1.13.1, Coin-OR branch and cut solver 2.9.5. [7]
2. Python 2.7 with Apache Thrift 0.8.0, Zope Interface 4.1.1, Twisted 14.0.2, NumPy 1.8.2, SciPy 0.14.1, Pandas 0.16.0, Statsmodesl 0.6.1, PuLP 1.6.0.
3. R 3.2.2 with Rcpp 0.12.1 ggplot2 1.0.1, grid 3.2.2, gridExtra 2.0.0, gtable 0.1.2, knitr 1.11, markdown 0.7.7, MASS 7.3-45, plyr 1.8.3, RColorBrewer 1.1-2, reshape 0.8.5, reshape2 1.4.1, rmarkdown 0.8.1, xtable 1.8-0, extrafont 0.17, tum[8] 1.0, and the Roboto font.[9]

4. Dynamic Time Warping[10] library and Flip-Flop filter library[11]

## 8. Revision comments

The research described in the paper is a great example of why making experiments reproducible is important, and how difficult this can be. The experimental setup used by authors is far from trivial as there are many details and configuration parameters involved. In fact, it can be very hard and time-consuming to replicate the results and re-use the described approach by simply reading the manuscript. Having a reproducible package facilitates this process.

The reproducibility report describes the simulation framework that authors implemented for their experiments; the same framework is used for both simulations and lab experiments in a real physical cloud infrastructure. In addition to the code (through GitHub) and data (time series with VMs timelines, also through GitHub), the authors created a ReproZip package and a Docker container, which makes it simple to reproduce the simulations without having to manually install all the required dependencies. These results could be successfully confirmed by reviewers.

We hope that the published assets will be of great help for readers interested in reproducing and re-using the experiments from the authors' original paper.

Corresponding author.

[1] Reviewer.

[2] The term domain is used as a synonym for VM within the source codes.

[3] https://github.com/jacksonicson/rain

[4] github.com/jacksonicson/times/tree/dcf54c84c166dc9693581a60101dc2ab2f2b92ea

[5] github.com/jacksonicson/paper.IS2015/tree/7165452f4e9c540f98e1e57058de06f9fb192e8f

[6] github.com/jacksonicson/workload/tree/713dc5382b82e4ec1e1b6a998c80af3f7c08219f

[7] https://projects.coin-or.org/Cbc

[8] github.com/jacksonicson/R.tum/tree/665fb85bc3ad259833b201218d735e688ffbb6c8

[9] https://www.google.de/design/spec/resources/roboto-noto-fonts.html

[10] github.com/jacksonicson/dtw/tree/1c9b0c53820fb7242e52870cf81bffaa7ae43681

[11] github.com/jacksonicson/fflop/tree/37fe72a663da6ffddbf2fc80bba0fdaf8fc52dcf